\documentclass[aps,prl,reprint,superscriptaddress,nofootinbib]{revtex4-2}

\usepackage[T1]{fontenc}
\usepackage[utf8]{inputenc}
\usepackage{amsmath,amssymb,bm}
\usepackage{graphicx}
\usepackage{siunitx}
\usepackage[colorlinks=true,allcolors=blue]{hyperref}

\newcommand{\diff}{\,\mathrm{d}}

\begin{document}

\typeout{COLUMNWIDTH=\the\columnwidth}
\typeout{TEXTWIDTH=\the\textwidth}

\title{A Free Sphere Reverses the Rebound Direction of a Near-Wall Cavitation Bubble}

\author{Chun-Zhu Ren}
\affiliation{School of Marine Science and Technology, Northwestern Polytechnical University, Xi'an, China}

\author{Jun Wen}
\affiliation{School of Marine Science and Technology, Northwestern Polytechnical University, Xi'an, China}

\author{Hai-Bao Hu}
\affiliation{School of Marine Science and Technology, Northwestern Polytechnical University, Xi'an, China}

\author{A-Man Zhang}
\affiliation{College of Shipbuilding Engineering, Harbin Engineering University, Harbin, China}

\author{Xiao Huang}
\email{huangxiao@nwpu.edu.cn}

\affiliation{School of Marine Science and Technology, Northwestern Polytechnical University, Xi'an, China}

\date{\today}

\begin{abstract}
A near-wall cavitation bubble is generally expected to acquire a wallward
Kelvin-impulse bias and to rebound or jet toward the wall. Here we
show that this canonical direction can be reversed by a wall-supported free
sphere. High-speed imaging reveals a transition
from away-from-wall to wallward rebound as the initial bubble--sphere separation
is increased.  By reconstructing the Kelvin impulse on a closed bubble boundary
that includes both the visible free interface and the bubble-side contact
closure, we find that the reversal is not governed primarily by the
instantaneous velocity of the sphere.  Instead, sphere displacement creates a
contact closure on which the bubble-source contribution supplies an
away-from-wall impulse.  This contact-source impulse competes with a wallward
background formed by the wall-image source and the quadrupolar component of the
sphere-induced field.  The resulting balance yields a calibrated geometric
criterion, \(\mathcal M_K\), and, in the comparable-size bubble--sphere regime,
reduces to a contact number \(a_z z_b/R_K^2\).  These results identify a
contact-geometric mechanism by which a movable particle can redirect the
first-cycle jet and rebound bias of a near-wall cavitation bubble.
\end{abstract}

\maketitle

Near a rigid wall, a collapsing cavitation bubble usually develops a wallward
Kelvin-impulse bias and a wall-directed jet
\cite{Rayleigh1917,Blake1986,BestBlake1994}.  This directionality concentrates
liquid inertia and impulsive loading on the solid surface and is a central route
to cavitation erosion \cite{Vogel1989,Philipp1998,Brujan2002,Lindau2003}.  A
major way of reducing this loading is therefore to alter the collapse asymmetry
or redirect the post-collapse motion.  Such redirection has been observed near
gas-entrapping surfaces \cite{GonzalezAvila2020,Sun2025JFMGasEntrap}, in
externally imposed flow fields \cite{Mnich2024JFMStagnation}, and near curved or
compound boundaries \cite{Tomita2002,Obreschkow2011,Supponen2016,Tagawa2018}.
These studies show that the rebound or jet direction is not merely a passive
consequence of bubble collapse, but can be selected by the surrounding
boundaries.

This direction-selection problem becomes more subtle when the boundary is a
particle that can move.  Cavitation bubbles can accelerate free particles and
modify particle-scale flows \cite{Poulain2015,Wu2017,Wu2021}, while recent
particle-laden configurations show that particles can also reshape cavitation
jetting and lift-off dynamics \cite{Cheng2025NatCommun,Ren2025JFMParticleJump}.
For a particle initially resting on a rigid substrate, a laser-induced bubble
can lift the particle from the wall \cite{Ren2022}.  However, lift-off alone
does not determine whether the particle subsequently remains separated from the
wall or is driven back toward it.  That later outcome is controlled by the
rebound or jet direction selected by the bubble near the end of the first
cycle.  A criterion for this direction in a movable-boundary geometry is still
lacking.

Here we show that a wall-supported free sphere can reverse the canonical
wallward rebound direction of a near-wall cavitation bubble within the first
bubble cycle.  High-speed imaging reveals a transition from away-from-wall to
wallward rebound as the initial bubble--sphere separation is increased.  To
identify the mechanism, we reconstruct the Kelvin impulse on a closed bubble
boundary composed of the visible free interface and the bubble-side contact
closure.  The reversal is not governed primarily by the instantaneous velocity
of the sphere.  Instead, the lifted sphere changes the closed boundary on which
the bubble Kelvin impulse is defined: the bubble-source contribution over the
contact closure supplies an away-from-wall impulse, competing with a wallward
background from the wall-image source and the quadrupolar sphere-induced field.
Their balance yields a geometric impulse criterion, $\mathcal{M}_K$, which
separates away-from-wall from wallward rebound and reduces, in the
comparable-size bubble--sphere regime, to the contact number $a_z z_b/R_K^2$.

\begin{figure}[t]
\centering
\includegraphics[width=0.8\columnwidth]{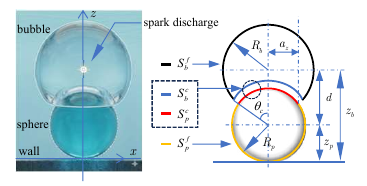}
\caption{
Experimental configuration and geometric notation. A spark-generated cavitation bubble interacts with a wall-supported free sphere near a rigid wall.
}

\label{fig:setup}
\end{figure}

\begin{figure*}[t]
\centering
\includegraphics[width=\textwidth]{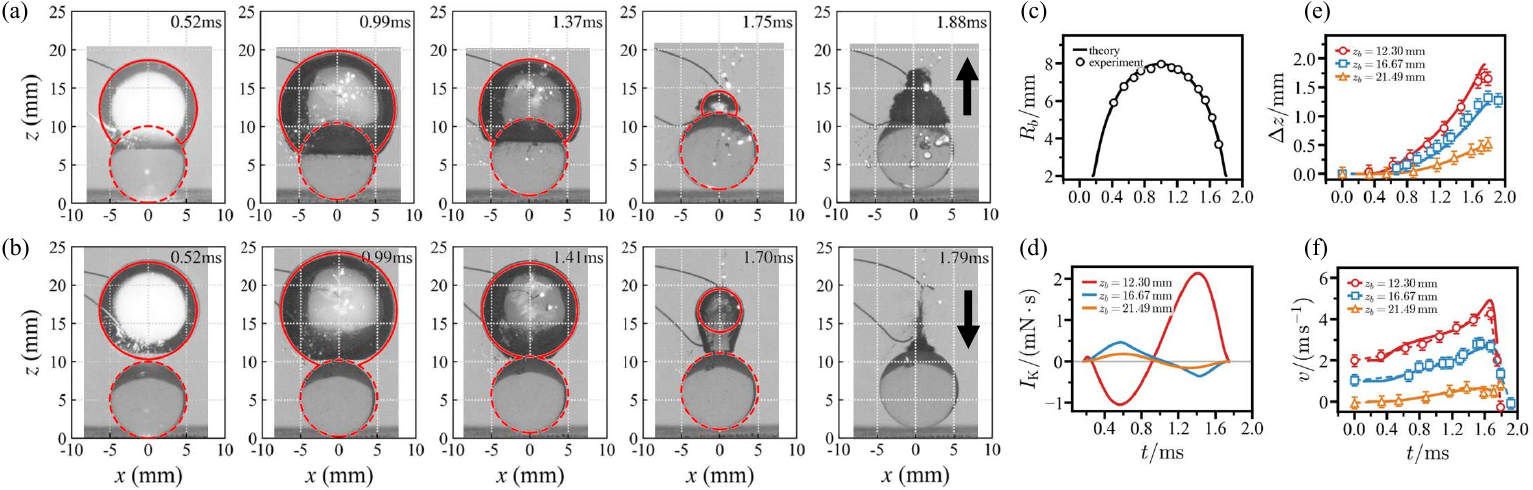}

\caption{
bubble-jet reversal and computed sphere motion for \(R_{b,\max}=7.56\,\mathrm{mm}\).
(a,b) Representative high-speed image sequences for \(z_b=12.30\,\mathrm{mm}\) and
\(z_b=16.67\,\mathrm{mm}\), showing away-from-wall and wallward rebound, respectively.
Arrows indicate the observed post-first-cycle rebound or jet direction; solid and dashed overlays denote the calculated bubble outline and sphere position.
(c--f) Measured radius history \(R_b(t)\), vertical Kelvin impulse \(I_K(t)\), sphere displacement \(\Delta z_p(t)\), and sphere velocity \(v_p(t)\).
Red, blue, and orange denote \(z_b=12.30\), \(16.67\), and \(21.49\,\mathrm{mm}\), respectively.
In (f), successive velocity curves are vertically offset by \(1.0\,\mathrm{m\,s^{-1}}\) for clarity.
}

\label{fig:snapshots}
\end{figure*}

The experimental configuration and notation are shown in Fig.~\ref{fig:setup}.
A hydrophilic glass sphere of radius $R_p=\SI{5}{mm}$ and density $\rho_p=\SI{2500}{kg.m^{-3}}$ initially rests on a rigid marble wall.
A single cavitation bubble is generated by controlled spark discharge on the
symmetry axis of the sphere at height $z_b$
\cite{Kling1972,Karri2012,Poulain2015,Teran2018,Wu2021,Ren2022,Huang2024PoF}.
The sphere-center position is denoted by $z_p(t)$, with $z_{p,0}=R_p$, so that the initial bubble--sphere center distance is $d_0=z_b-z_{p,0}$.
Experiments are performed for different maximum bubble radii $R_{b,\max}$ and initial distances $d_0$.
The bubble-radius history $R_b(t)$, the sphere trajectory $z_p(t)$, and the rebound direction are extracted from side-view high-speed images recorded by a Phantom camera at $24000$ frames per second with an exposure time of $\SI{1.5}{\micro\second}$.
The measured \(R_b(t)\) corresponds to the envelope of the visible free-interface portion \(S_b^f\) of the bubble.
For the Kelvin-impulse evaluation, the closed bubble boundary is completed by adding the bubble-side contact surface \(S_b^c\) when contact occurs.

Figures~\ref{fig:snapshots}(a) and \ref{fig:snapshots}(b) show two representative bubble--sphere interactions with opposite post-first-cycle jet or rebound biases.  At the smaller separation, $z_b=12.30,\mathrm{mm}$, strong near-field contact lifts the sphere from the wall and the bubble rebounds away from the wall.  At the larger separation, $z_b=16.67,\mathrm{mm}$, the bubble remains wallward biased despite the presence of the sphere.

To quantify this transition, we compute the coupled flow and sphere motion during the first bubble cycle.  The liquid is treated as inviscid, incompressible, and irrotational, with instantaneous velocity potential $\phi(\mathbf{x},t;Q,z_b,z_p,v_p)$ \cite{Lamb1932,MilneThomson1968,Brennen2014}.  The measured radius history $R_b(t)$ prescribes the bubble-source strength $Q(t)=4\pi R_b^2\dot R_b$ \cite{Rayleigh1917,Leighton1994,Zhang2023Unified}; the rigid wall is imposed by the method of images \cite{Blake1986,BestBlake1994,Blake2015}; and the sphere-induced correction is represented by a multipole expansion satisfying no penetration on the moving sphere \cite{Weiss1944,Lamb1932,LandweberMiloh1980,BestBlake1994,Wang2022Weiss}.  The sphere motion is obtained from a partitioned pressure calculation: the liquid-wetted surface uses the unsteady Bernoulli pressure $p_f=\rho(\partial_t\phi-|\nabla\phi|^2/2)$, while the forced bubble--sphere contact cap is assigned a power-consistent mean pressure.

The measured $R_b(t)$ used as the source input is shown in Fig.~\ref{fig:snapshots}(c).  With this input, the computed sphere displacement and velocity reproduce the observed lift-off, acceleration, and deceleration trends, as shown in Figs.~\ref{fig:snapshots}(e) and \ref{fig:snapshots}(f).  We then reconstruct the bubble Kelvin impulse on the closed bubble boundary, composed of the visible free-interface part $S_b^f$ and the bubble-side contact closure $S_b^c$:

\begin{equation}
\mathbf{I}_K = 
-\rho
\left[
\int_{S_b^f}\phi_b^f\,\mathbf{n}\,\diff S
+
\int_{S_b^c}\phi_b^c\,\mathbf{n}\,\diff S
\right].
\label{eq:IK_closed}
\end{equation}

Here $\phi_b^{f}$ is obtained from the analytical potential-flow model on $S_b^f$.  On the forced contact cap $S_b^c$, the contact potential $\phi_b^{c}$ is reconstructed from the imposed contact geometry and boundary conditions using a boundary-integral procedure \cite{Blake1986,Zhang1993,BestBlake1994}.  The resulting vertical impulse $I_K(t)$, shown in Fig.~\ref{fig:snapshots}(d), changes sign consistently with the observed jet or rebound bias: the away-from-wall case develops an upward-biased impulse, whereas the wallward case retains a downward-biased impulse.  We therefore decompose the reconstructed Kelvin impulse to identify the boundary contribution responsible for the direction reversal.

We then ask which part of the closed boundary and which potential component dominate this impulse bias. To this end, the vertical Kelvin impulse is decomposed into the bubble-source contribution, wall-image-source contribution, sphere-induced contribution, image-sphere contribution, and moving-contact velocity contributions, denoted by $I_{\rm bub}$, $I_{\rm ibub}$, $I_{\rm sph}$, $I_{\rm isph}$, and $I_{\rm vel}$, respectively:
\begin{equation}
I_{K}
\simeq
I_{\mathrm{bub}}
+
I_{\mathrm{ibub}}
+
I_{\mathrm{sph}}
+
I_{\mathrm{isph}}
+
I_{\mathrm{vel}} .
\label{eq:impulse_decomp}
\end{equation}
This decomposition is used below to identify the dominant term responsible for the reversal of the bubble impulse.

\begin{figure*}[t]
\centering
\includegraphics[width=\textwidth]{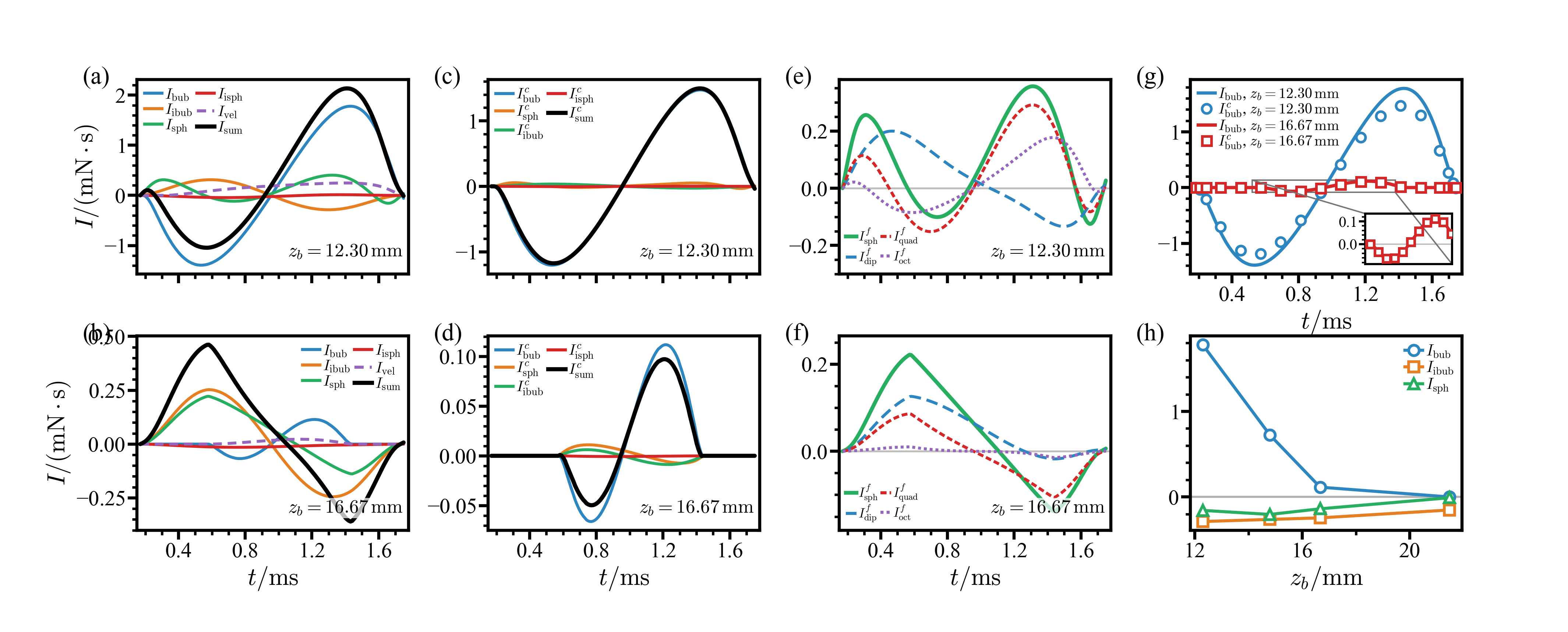}
\caption{
Decomposition of the reconstructed first-cycle Kelvin impulse.
(a,b) Closed-boundary decomposition for \(z_b=12.30\,\mathrm{mm}\) and
\(z_b=16.67\,\mathrm{mm}\).  The subscripts bub, ibub, sph, isph, and vel
denote the bubble-source, wall-image-source, sphere-induced, image-sphere, and
velocity-boundary contributions, respectively.
(c,d) Corresponding decomposition on the contact closure surface, where the
bubble-source term \(I_{\rm bub}^{c}\) dominates.
(e,f) Free-interface decomposition of the wall-image contribution
\(I_{\rm ibub}^{f}\) and the multipolar sphere-induced contributions; the
near-end-cycle downward signature is mainly quadrupolar,
\(I_{\rm quad}^{f}\).
(g) Comparison of \(I_{\rm bub}\) and \(I_{\rm bub}^{c}\).
(h) Late first-cycle contributions versus \(z_b\), showing the weak sensitivity
of the negative wall--sphere background relative to \(I_{\rm bub}\).
Superscripts \(c\) and \(f\) denote contact-surface and free-interface
contributions.
}
\label{fig:impulse_decomp}
\end{figure*}

Figure~\ref{fig:impulse_decomp} identifies the dominant contributions to the
impulse bias.  On the closed bubble boundary, \(I_{\rm bub}\) supplies the
positive, away-from-wall impulse, whereas \(I_{\rm ibub}\) and \(I_{\rm sph}\)
provide the negative wallward contribution; \(I_{\rm isph}\) and \(I_{\rm vel}\)
remain secondary, as shown in Figs.~\ref{fig:impulse_decomp}(a,b).  This is consistent
with the conclusion that the sphere velocity gives only a finite correction,
whereas the dominant variation comes from the bubble-source contribution over
the contact part, controlled by the first-cycle contact area.  Restricting the
integral to the contact closure surface confirms that the contact impulse is
almost entirely \(I_{\rm bub}^{c}\)
Figs.~\ref{fig:impulse_decomp}(c,d), and the total bubble-source impulse
closely follows it, \(I_{\rm bub}\simeq I_{\rm bub}^{c}\)
[Fig.~\ref{fig:impulse_decomp}(g)].  On the free interface, the near-end-cycle
downward signature of the sphere-induced contribution is mainly captured by the
quadrupolar component, \(I_{\rm quad}^{f}\), while the wall-image source
contributes \(I_{\rm ibub}^{f}\), as shown in Figs.~\ref{fig:impulse_decomp}(e,f).  The
\(z_b\)-sweep further serves as a sensitivity check: \(I_{\rm ibub}\) and
\(I_{\rm sph}\) vary much more weakly with \(z_b\) than \(I_{\rm bub}\)
Fig.~\ref{fig:impulse_decomp}(h).  Thus, for the leading-order scaling, the
negative wallward impulse can be treated as a relatively stable background
threshold, and the reversal problem reduces to a competition between the
contact-source impulse \(I_{\rm bub}^{c}\) and the free-interface contributions
from the wall-image source \(I_{\rm ibub}^{f}\) and the quadrupolar
sphere-induced field \(I_{\rm quad}^{f}\).

Motivated by the decomposition above, we next validate the three dominant impulse scalings. Using the contact-pressure treatment for the forced contact cap \cite{Borkent2008,Li2018PoF,Ren2022}, the standard wall-image construction \cite{Blake1986,BestBlake1994,Blake2015}, and the quadrupole formulation for a spherical boundary \cite{Weiss1944,LandweberMiloh1980,BestBlake1994,Wang2022Weiss}, the corresponding scaling bases are $-\rho Q_b a_z^2/R_b$, $\rho Q_bR_b^3/z_b^2$, and $\rho Q_bR_p\mathcal{T}(R_b/R_p,d/R_p,\alpha)$, respectively.  Here $d=|z_b-z_p|$, $\alpha$ is the opening angle of the bubble free interface, and $\mathcal T$ is the dimensionless kernel obtained from the analytic free-interface quadrupole integral, whose explicit form is given in the Supplemental Material.  The collapses in Figs.~\ref{fig:criterion}(a--c) validate these scalings and determine the component-level coefficients $C_c$, $C_w$, and $C_{\rm quad}$.

The quadrupolar contribution is then evaluated at $t_K$, chosen as the instant at which $Q_b$ is most negative.  In the present data set, $0.495\le R_K/R_p\le 1.632$ and $0.880\le d_K/R_p\le 3.464$.  Over this range, the finite-size correction proportional to $(1-R_K/R_p)^2$ is secondary to the dominant $(d_K/R_p)^2$ dependence, so that the late-cycle kernel reduces to $\mathcal{T}_K\simeq -(1/32)(R_p/d_K)^3[2R_K/R_p-(d_K/R_p)^2]$. For non-contact cases, $a_z=0$ and the free interface is complete, so that
$\alpha=\pi$ and the quadrupolar kernel reduces to
$R_b^3R_p^4/d^7$. These cases therefore contain no
away-from-wall contact-source contribution, and the quadrupolar term only
enters as a correction to the wallward background.

\begin{figure*}[t]
\centering
\includegraphics[width=\textwidth]{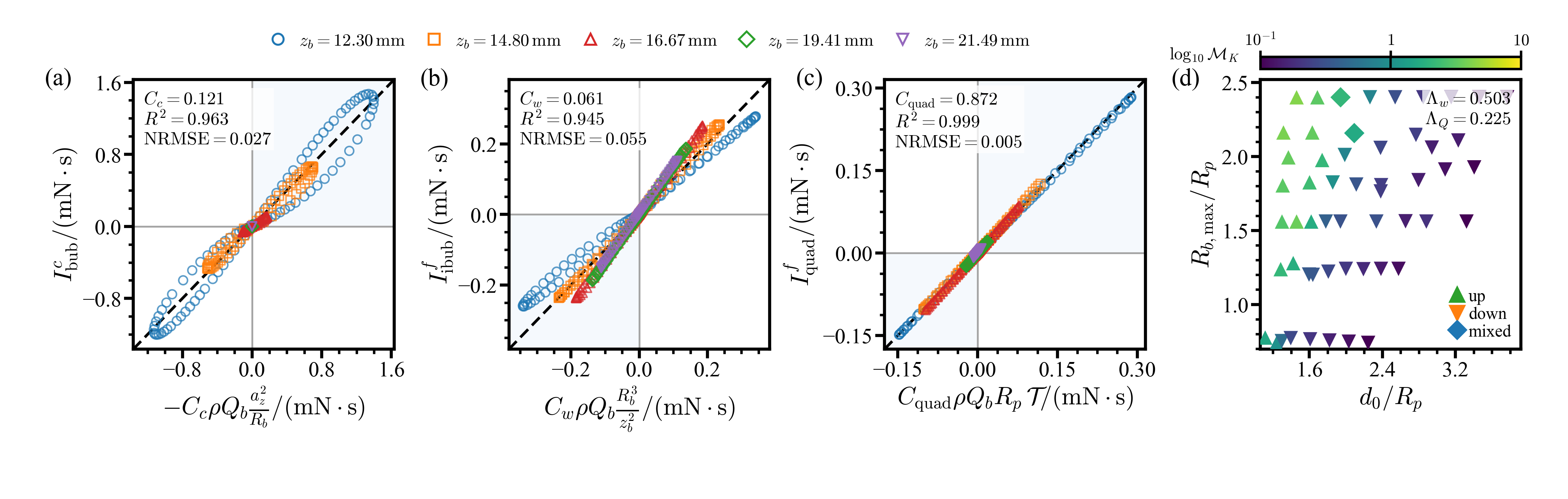}
\caption{
Calibration and test of the impulse criterion.
(a--c) BEM contributions versus the corresponding scaling predictions for
the contact bubble-source, wall-image-source, and quadrupolar sphere-induced
terms.  The dashed lines denote one-to-one agreement. (d) Rebound-direction classification by the calibrated geometric criterion
\(\mathcal{M}_K\).  The conditions \(\mathcal{M}_K>1\) and
\(\mathcal{M}_K<1\) predict away-from-wall and wallward rebound, respectively.
Colors denote \(\log_{10}\mathcal{M}_K\), with
\(\log_{10}\mathcal{M}_K=0\) marking the threshold \(\mathcal{M}_K=1\).
}

\label{fig:criterion}
\end{figure*}

Combining the validated component scalings gives the late-cycle vertical impulse balance
\begin{equation}
\begin{aligned}
I_K
&\simeq
-\frac{\rho Q_K C_c}{R_K}
\left[
a_z^2
+2\Lambda_Q\frac{R_K^2R_p^3}{d_K^3}
-\Lambda_Q\frac{R_KR_p^2}{d_K}
-\Lambda_w\frac{R_K^4}{z_b^2}
\right],
\end{aligned}
\label{eq:IK_balance_compact}
\end{equation}
where $R_K=R_b(t_K)$, $d_K=d(t_K)$, and $Q_K=Q_b(t_K)<0$.  The two weights are not fitted to the observed rebound directions.  They are fixed by the component-level coefficients in Figs.~\ref{fig:criterion}(a--c):
\begin{equation}
\Lambda_w=\left|\frac{C_w}{C_c}\right|,
\qquad
\Lambda_Q=\left|\frac{C_{\rm quad}}{32C_c}\right|.
\label{eq:Lambda_def}
\end{equation}
The factor $1/32$ in $\Lambda_Q$ comes from the analytical quadrupolar reduction.  Since the prefactor $-\rho Q_KC_c/R_K$ is positive, the rebound direction is determined by the bracketed geometric balance.  Using $R_p$ to nondimensionalize lengths, we define
\begin{equation}
\begin{aligned}
\mathcal{M}_K
&=
\frac{
\hat a_z^2+2\Lambda_Q\hat R_K^2\hat d_K^{-3}
}{
\Lambda_Q\hat R_K\hat d_K^{-1}
+\Lambda_w\hat R_K^4\hat z_b^{-2}
}.
\end{aligned}
\label{eq:MK}
\end{equation}

Here $\hat a_z=a_z/R_p$, $\hat R_K=R_K/R_p$, $\hat d_K=d_K/R_p$, and $\hat z_b=z_b/R_p$.  The criterion $\mathcal{M}_K>1$ predicts an away-from-wall impulse, whereas $\mathcal{M}_K<1$ predicts a wallward impulse.  Thus the classification in Fig.~\ref{fig:criterion}(d) is obtained without any direct fitting to the rebound-direction labels.

In the present parameter range, the quadrupolar term is comparable to the
wall-image term but is far less sensitive to geometry than the contact-source
term. The full balance is quantified by $\mathcal{M}_K$ and tested in
Fig.~\ref{fig:criterion}(d). Absorbing the quadrupolar correction into an
effective wallward background gives the reduced contact criterion
\begin{equation}
\frac{a_z z_b}{R_K^2}=C_* .
\label{eq:simple_criterion}
\end{equation}
For the present data set, $C_*\simeq 2.7$; larger values predict
away-from-wall rebound and smaller values predict wallward rebound. This
reduction shows that the reversal is controlled primarily by whether the
projected contact-source impulse exceeds the effective wallward background.

In conclusion, we have shown that a wall-supported free sphere can reverse the first-cycle Kelvin-impulse bias of a near-wall cavitation bubble.  The mechanism is not governed primarily by the instantaneous velocity of the sphere.  Instead, in the comparable-size bubble--sphere regime studied here, the upward impulse is supplied by the bubble-source contribution over the contact part of the closed bubble boundary, with its strength set by the projected contact scale $a_z$.  The opposing wallward impulse is supplied by two background contributions: the wall-image source and the quadrupolar component of the sphere-induced field.

This balance is first captured by the full impulse criterion $\mathcal{M}_K$, which retains the contact-source, wall-image, and quadrupolar sphere-induced terms.  Because the wall-image and quadrupolar contributions vary much less sensitively than the contact-source term over the present data set, they can be absorbed into an effective wallward background, yielding the reduced contact number $a_z z_b/R_K^2$.  The two-level criterion therefore identifies the same physical transition: rebound away from the wall occurs when the contact-source impulse exceeds the wallward background, whereas wallward rebound persists when the background dominates.

These results show that a surface sphere is not merely accelerated by a cavitation bubble.  Once displaced, it modifies the closed boundary on which the bubble Kelvin impulse is defined and thereby redirects the bubble's jet and rebound bias within the first cycle.  This contact-geometric mechanism provides a route for predicting whether a lifted sphere remains separated from the wall as an energy-absorbing movable boundary or returns toward the wall as a potential secondary impact source for cavitation damage.

\begin{acknowledgments}
The authors acknowledge financial support from the National Natural Science
Foundation of China (No. 52471345) and the Innovation Capability Support
Program of Shaanxi (No. 2024RS-CXTD-15).
\end{acknowledgments}

\bibliographystyle{apsrev4-2}
\bibliography{references}

\begin{thebibliography}{36}%
\makeatletter
\providecommand \@ifxundefined [1]{%
 \@ifx{#1\undefined}
}%
\providecommand \@ifnum [1]{%
 \ifnum #1\expandafter \@firstoftwo
 \else \expandafter \@secondoftwo
 \fi
}%
\providecommand \@ifx [1]{%
 \ifx #1\expandafter \@firstoftwo
 \else \expandafter \@secondoftwo
 \fi
}%
\providecommand \natexlab [1]{#1}%
\providecommand \enquote  [1]{``#1''}%
\providecommand \bibnamefont  [1]{#1}%
\providecommand \bibfnamefont [1]{#1}%
\providecommand \citenamefont [1]{#1}%
\providecommand \href@noop [0]{\@secondoftwo}%
\providecommand \href [0]{\begingroup \@sanitize@url \@href}%
\providecommand \@href[1]{\@@startlink{#1}\@@href}%
\providecommand \@@href[1]{\endgroup#1\@@endlink}%
\providecommand \@sanitize@url [0]{\catcode `\\12\catcode `\$12\catcode `\&12\catcode `\#12\catcode `\^12\catcode `\_12\catcode `\%12\relax}%
\providecommand \@@startlink[1]{}%
\providecommand \@@endlink[0]{}%
\providecommand \url  [0]{\begingroup\@sanitize@url \@url }%
\providecommand \@url [1]{\endgroup\@href {#1}{\urlprefix }}%
\providecommand \urlprefix  [0]{URL }%
\providecommand \Eprint [0]{\href }%
\providecommand \doibase [0]{https://doi.org/}%
\providecommand \selectlanguage [0]{\@gobble}%
\providecommand \bibinfo  [0]{\@secondoftwo}%
\providecommand \bibfield  [0]{\@secondoftwo}%
\providecommand \translation [1]{[#1]}%
\providecommand \BibitemOpen [0]{}%
\providecommand \bibitemStop [0]{}%
\providecommand \bibitemNoStop [0]{.\EOS\space}%
\providecommand \EOS [0]{\spacefactor3000\relax}%
\providecommand \BibitemShut  [1]{\csname bibitem#1\endcsname}%
\let\auto@bib@innerbib\@empty
\bibitem [{\citenamefont {Rayleigh}(1917)}]{Rayleigh1917}%
  \BibitemOpen
  \bibfield  {author} {\bibinfo {author} {\bibfnamefont {L.}~\bibnamefont {Rayleigh}},\ }\href {https://doi.org/10.1080/14786440808635681} {\bibfield  {journal} {\bibinfo  {journal} {The London, Edinburgh, and Dublin Philosophical Magazine and Journal of Science}\ }\bibinfo {series} {6},\ \textbf {\bibinfo {volume} {34}},\ \bibinfo {pages} {94} (\bibinfo {year} {1917})}\BibitemShut {NoStop}%
\bibitem [{\citenamefont {Blake}\ \emph {et~al.}(1986)\citenamefont {Blake}, \citenamefont {Taib},\ and\ \citenamefont {Doherty}}]{Blake1986}%
  \BibitemOpen
  \bibfield  {author} {\bibinfo {author} {\bibfnamefont {J.~R.}\ \bibnamefont {Blake}}, \bibinfo {author} {\bibfnamefont {B.~B.}\ \bibnamefont {Taib}},\ and\ \bibinfo {author} {\bibfnamefont {G.}~\bibnamefont {Doherty}},\ }\href {https://doi.org/10.1017/S0022112086000988} {\bibfield  {journal} {\bibinfo  {journal} {Journal of Fluid Mechanics}\ }\textbf {\bibinfo {volume} {170}},\ \bibinfo {pages} {479} (\bibinfo {year} {1986})}\BibitemShut {NoStop}%
\bibitem [{\citenamefont {Best}\ and\ \citenamefont {Blake}(1994)}]{BestBlake1994}%
  \BibitemOpen
  \bibfield  {author} {\bibinfo {author} {\bibfnamefont {J.~P.}\ \bibnamefont {Best}}\ and\ \bibinfo {author} {\bibfnamefont {J.~R.}\ \bibnamefont {Blake}},\ }\href {https://doi.org/10.1017/S0022112094000273} {\bibfield  {journal} {\bibinfo  {journal} {Journal of Fluid Mechanics}\ }\textbf {\bibinfo {volume} {261}},\ \bibinfo {pages} {75} (\bibinfo {year} {1994})}\BibitemShut {NoStop}%
\bibitem [{\citenamefont {Vogel}\ \emph {et~al.}(1989)\citenamefont {Vogel}, \citenamefont {Lauterborn},\ and\ \citenamefont {Timm}}]{Vogel1989}%
  \BibitemOpen
  \bibfield  {author} {\bibinfo {author} {\bibfnamefont {A.}~\bibnamefont {Vogel}}, \bibinfo {author} {\bibfnamefont {W.}~\bibnamefont {Lauterborn}},\ and\ \bibinfo {author} {\bibfnamefont {R.}~\bibnamefont {Timm}},\ }\href {https://doi.org/10.1017/S0022112089002314} {\bibfield  {journal} {\bibinfo  {journal} {Journal of Fluid Mechanics}\ }\textbf {\bibinfo {volume} {206}},\ \bibinfo {pages} {299} (\bibinfo {year} {1989})}\BibitemShut {NoStop}%
\bibitem [{\citenamefont {Philipp}\ and\ \citenamefont {Lauterborn}(1998)}]{Philipp1998}%
  \BibitemOpen
  \bibfield  {author} {\bibinfo {author} {\bibfnamefont {A.}~\bibnamefont {Philipp}}\ and\ \bibinfo {author} {\bibfnamefont {W.}~\bibnamefont {Lauterborn}},\ }\href {https://doi.org/10.1017/S0022112098008738} {\bibfield  {journal} {\bibinfo  {journal} {Journal of Fluid Mechanics}\ }\textbf {\bibinfo {volume} {361}},\ \bibinfo {pages} {75} (\bibinfo {year} {1998})}\BibitemShut {NoStop}%
\bibitem [{\citenamefont {Brujan}\ \emph {et~al.}(2002)\citenamefont {Brujan}, \citenamefont {Keen}, \citenamefont {Vogel},\ and\ \citenamefont {Blake}}]{Brujan2002}%
  \BibitemOpen
  \bibfield  {author} {\bibinfo {author} {\bibfnamefont {E.~A.}\ \bibnamefont {Brujan}}, \bibinfo {author} {\bibfnamefont {G.~S.}\ \bibnamefont {Keen}}, \bibinfo {author} {\bibfnamefont {A.}~\bibnamefont {Vogel}},\ and\ \bibinfo {author} {\bibfnamefont {J.~R.}\ \bibnamefont {Blake}},\ }\href {https://doi.org/10.1063/1.1421102} {\bibfield  {journal} {\bibinfo  {journal} {Physics of Fluids}\ }\textbf {\bibinfo {volume} {14}},\ \bibinfo {pages} {85} (\bibinfo {year} {2002})}\BibitemShut {NoStop}%
\bibitem [{\citenamefont {Lindau}\ and\ \citenamefont {Lauterborn}(2003)}]{Lindau2003}%
  \BibitemOpen
  \bibfield  {author} {\bibinfo {author} {\bibfnamefont {O.}~\bibnamefont {Lindau}}\ and\ \bibinfo {author} {\bibfnamefont {W.}~\bibnamefont {Lauterborn}},\ }\href {https://doi.org/10.1017/S0022112002003695} {\bibfield  {journal} {\bibinfo  {journal} {Journal of Fluid Mechanics}\ }\textbf {\bibinfo {volume} {479}},\ \bibinfo {pages} {327} (\bibinfo {year} {2003})}\BibitemShut {NoStop}%
\bibitem [{\citenamefont {Gonzalez-Avila}\ \emph {et~al.}(2020)\citenamefont {Gonzalez-Avila}, \citenamefont {Nguyen}, \citenamefont {Arunachalam}, \citenamefont {Domingues}, \citenamefont {Mishra},\ and\ \citenamefont {Ohl}}]{GonzalezAvila2020}%
  \BibitemOpen
  \bibfield  {author} {\bibinfo {author} {\bibfnamefont {S.~R.}\ \bibnamefont {Gonzalez-Avila}}, \bibinfo {author} {\bibfnamefont {D.~M.}\ \bibnamefont {Nguyen}}, \bibinfo {author} {\bibfnamefont {S.}~\bibnamefont {Arunachalam}}, \bibinfo {author} {\bibfnamefont {E.~M.}\ \bibnamefont {Domingues}}, \bibinfo {author} {\bibfnamefont {H.}~\bibnamefont {Mishra}},\ and\ \bibinfo {author} {\bibfnamefont {C.-D.}\ \bibnamefont {Ohl}},\ }\href {https://doi.org/10.1126/sciadv.aax6192} {\bibfield  {journal} {\bibinfo  {journal} {Science Advances}\ }\textbf {\bibinfo {volume} {6}},\ \bibinfo {pages} {eaax6192} (\bibinfo {year} {2020})}\BibitemShut {NoStop}%
\bibitem [{\citenamefont {Sun}\ \emph {et~al.}(2025)\citenamefont {Sun}, \citenamefont {Yao}, \citenamefont {Wang}, \citenamefont {Zhong}, \citenamefont {Xiao}, \citenamefont {Ohl},\ and\ \citenamefont {Wang}}]{Sun2025JFMGasEntrap}%
  \BibitemOpen
  \bibfield  {author} {\bibinfo {author} {\bibfnamefont {Y.}~\bibnamefont {Sun}}, \bibinfo {author} {\bibfnamefont {Z.}~\bibnamefont {Yao}}, \bibinfo {author} {\bibfnamefont {C.}~\bibnamefont {Wang}}, \bibinfo {author} {\bibfnamefont {Q.}~\bibnamefont {Zhong}}, \bibinfo {author} {\bibfnamefont {R.}~\bibnamefont {Xiao}}, \bibinfo {author} {\bibfnamefont {C.-D.}\ \bibnamefont {Ohl}},\ and\ \bibinfo {author} {\bibfnamefont {F.}~\bibnamefont {Wang}},\ }\href {https://doi.org/10.1017/jfm.2025.399} {\bibfield  {journal} {\bibinfo  {journal} {Journal of Fluid Mechanics}\ }\textbf {\bibinfo {volume} {1013}},\ \bibinfo {pages} {A9} (\bibinfo {year} {2025})}\BibitemShut {NoStop}%
\bibitem [{\citenamefont {Mnich}\ \emph {et~al.}(2024)\citenamefont {Mnich}, \citenamefont {Reuter}, \citenamefont {Denner},\ and\ \citenamefont {Ohl}}]{Mnich2024JFMStagnation}%
  \BibitemOpen
  \bibfield  {author} {\bibinfo {author} {\bibfnamefont {D.}~\bibnamefont {Mnich}}, \bibinfo {author} {\bibfnamefont {F.}~\bibnamefont {Reuter}}, \bibinfo {author} {\bibfnamefont {F.}~\bibnamefont {Denner}},\ and\ \bibinfo {author} {\bibfnamefont {C.-D.}\ \bibnamefont {Ohl}},\ }\href {https://doi.org/10.1017/jfm.2023.1048} {\bibfield  {journal} {\bibinfo  {journal} {Journal of Fluid Mechanics}\ }\textbf {\bibinfo {volume} {979}},\ \bibinfo {pages} {A18} (\bibinfo {year} {2024})}\BibitemShut {NoStop}%
\bibitem [{\citenamefont {Tomita}\ \emph {et~al.}(2002)\citenamefont {Tomita}, \citenamefont {Robinson}, \citenamefont {Tong},\ and\ \citenamefont {Blake}}]{Tomita2002}%
  \BibitemOpen
  \bibfield  {author} {\bibinfo {author} {\bibfnamefont {Y.}~\bibnamefont {Tomita}}, \bibinfo {author} {\bibfnamefont {P.~B.}\ \bibnamefont {Robinson}}, \bibinfo {author} {\bibfnamefont {R.~P.}\ \bibnamefont {Tong}},\ and\ \bibinfo {author} {\bibfnamefont {J.~R.}\ \bibnamefont {Blake}},\ }\href {https://doi.org/10.1017/S0022112002001209} {\bibfield  {journal} {\bibinfo  {journal} {Journal of Fluid Mechanics}\ }\textbf {\bibinfo {volume} {466}},\ \bibinfo {pages} {259} (\bibinfo {year} {2002})}\BibitemShut {NoStop}%
\bibitem [{\citenamefont {Obreschkow}\ \emph {et~al.}(2011)\citenamefont {Obreschkow}, \citenamefont {Tinguely}, \citenamefont {Dorsaz}, \citenamefont {Kobel}, \citenamefont {de~Bosset},\ and\ \citenamefont {Farhat}}]{Obreschkow2011}%
  \BibitemOpen
  \bibfield  {author} {\bibinfo {author} {\bibfnamefont {D.}~\bibnamefont {Obreschkow}}, \bibinfo {author} {\bibfnamefont {M.}~\bibnamefont {Tinguely}}, \bibinfo {author} {\bibfnamefont {N.}~\bibnamefont {Dorsaz}}, \bibinfo {author} {\bibfnamefont {P.}~\bibnamefont {Kobel}}, \bibinfo {author} {\bibfnamefont {A.}~\bibnamefont {de~Bosset}},\ and\ \bibinfo {author} {\bibfnamefont {M.}~\bibnamefont {Farhat}},\ }\href {https://doi.org/10.1103/PhysRevLett.107.204501} {\bibfield  {journal} {\bibinfo  {journal} {Physical Review Letters}\ }\textbf {\bibinfo {volume} {107}},\ \bibinfo {pages} {204501} (\bibinfo {year} {2011})}\BibitemShut {NoStop}%
\bibitem [{\citenamefont {Supponen}\ \emph {et~al.}(2016)\citenamefont {Supponen}, \citenamefont {Obreschkow}, \citenamefont {Tinguely}, \citenamefont {Kobel}, \citenamefont {Dorsaz},\ and\ \citenamefont {Farhat}}]{Supponen2016}%
  \BibitemOpen
  \bibfield  {author} {\bibinfo {author} {\bibfnamefont {O.}~\bibnamefont {Supponen}}, \bibinfo {author} {\bibfnamefont {D.}~\bibnamefont {Obreschkow}}, \bibinfo {author} {\bibfnamefont {M.}~\bibnamefont {Tinguely}}, \bibinfo {author} {\bibfnamefont {P.}~\bibnamefont {Kobel}}, \bibinfo {author} {\bibfnamefont {N.}~\bibnamefont {Dorsaz}},\ and\ \bibinfo {author} {\bibfnamefont {M.}~\bibnamefont {Farhat}},\ }\href {https://doi.org/10.1017/jfm.2016.463} {\bibfield  {journal} {\bibinfo  {journal} {Journal of Fluid Mechanics}\ }\textbf {\bibinfo {volume} {802}},\ \bibinfo {pages} {263} (\bibinfo {year} {2016})}\BibitemShut {NoStop}%
\bibitem [{\citenamefont {Tagawa}\ and\ \citenamefont {Peters}(2018)}]{Tagawa2018}%
  \BibitemOpen
  \bibfield  {author} {\bibinfo {author} {\bibfnamefont {Y.}~\bibnamefont {Tagawa}}\ and\ \bibinfo {author} {\bibfnamefont {I.~R.}\ \bibnamefont {Peters}},\ }\href {https://doi.org/10.1103/PhysRevFluids.3.081601} {\bibfield  {journal} {\bibinfo  {journal} {Physical Review Fluids}\ }\textbf {\bibinfo {volume} {3}},\ \bibinfo {pages} {081601} (\bibinfo {year} {2018})}\BibitemShut {NoStop}%
\bibitem [{\citenamefont {Poulain}\ \emph {et~al.}(2015)\citenamefont {Poulain}, \citenamefont {Guenoun}, \citenamefont {Gart}, \citenamefont {Crowe},\ and\ \citenamefont {Jung}}]{Poulain2015}%
  \BibitemOpen
  \bibfield  {author} {\bibinfo {author} {\bibfnamefont {S.}~\bibnamefont {Poulain}}, \bibinfo {author} {\bibfnamefont {G.}~\bibnamefont {Guenoun}}, \bibinfo {author} {\bibfnamefont {S.}~\bibnamefont {Gart}}, \bibinfo {author} {\bibfnamefont {W.}~\bibnamefont {Crowe}},\ and\ \bibinfo {author} {\bibfnamefont {S.}~\bibnamefont {Jung}},\ }\href {https://doi.org/10.1103/PhysRevLett.114.214501} {\bibfield  {journal} {\bibinfo  {journal} {Physical Review Letters}\ }\textbf {\bibinfo {volume} {114}},\ \bibinfo {pages} {214501} (\bibinfo {year} {2015})}\BibitemShut {NoStop}%
\bibitem [{\citenamefont {Wu}\ \emph {et~al.}(2017)\citenamefont {Wu}, \citenamefont {Zuo}, \citenamefont {Stone},\ and\ \citenamefont {Liu}}]{Wu2017}%
  \BibitemOpen
  \bibfield  {author} {\bibinfo {author} {\bibfnamefont {S.}~\bibnamefont {Wu}}, \bibinfo {author} {\bibfnamefont {Z.}~\bibnamefont {Zuo}}, \bibinfo {author} {\bibfnamefont {H.~A.}\ \bibnamefont {Stone}},\ and\ \bibinfo {author} {\bibfnamefont {S.}~\bibnamefont {Liu}},\ }\href {https://doi.org/10.1103/PhysRevLett.119.084501} {\bibfield  {journal} {\bibinfo  {journal} {Physical Review Letters}\ }\textbf {\bibinfo {volume} {119}},\ \bibinfo {pages} {084501} (\bibinfo {year} {2017})}\BibitemShut {NoStop}%
\bibitem [{\citenamefont {Wu}\ \emph {et~al.}(2021)\citenamefont {Wu}, \citenamefont {Li}, \citenamefont {Zuo}, \citenamefont {Stone},\ and\ \citenamefont {Liu}}]{Wu2021}%
  \BibitemOpen
  \bibfield  {author} {\bibinfo {author} {\bibfnamefont {S.}~\bibnamefont {Wu}}, \bibinfo {author} {\bibfnamefont {B.}~\bibnamefont {Li}}, \bibinfo {author} {\bibfnamefont {Z.}~\bibnamefont {Zuo}}, \bibinfo {author} {\bibfnamefont {H.~A.}\ \bibnamefont {Stone}},\ and\ \bibinfo {author} {\bibfnamefont {S.}~\bibnamefont {Liu}},\ }\href {https://doi.org/10.1103/PhysRevFluids.6.093602} {\bibfield  {journal} {\bibinfo  {journal} {Physical Review Fluids}\ }\textbf {\bibinfo {volume} {6}},\ \bibinfo {pages} {093602} (\bibinfo {year} {2021})}\BibitemShut {NoStop}%
\bibitem [{\citenamefont {Cheng}\ \emph {et~al.}(2025)\citenamefont {Cheng}, \citenamefont {Chen}, \citenamefont {Yuan},\ and\ \citenamefont {Jia}}]{Cheng2025NatCommun}%
  \BibitemOpen
  \bibfield  {author} {\bibinfo {author} {\bibfnamefont {X.}~\bibnamefont {Cheng}}, \bibinfo {author} {\bibfnamefont {X.-P.}\ \bibnamefont {Chen}}, \bibinfo {author} {\bibfnamefont {Z.-M.}\ \bibnamefont {Yuan}},\ and\ \bibinfo {author} {\bibfnamefont {L.}~\bibnamefont {Jia}},\ }\href {https://doi.org/10.1038/s41467-025-62936-y} {\bibfield  {journal} {\bibinfo  {journal} {Nature Communications}\ }\textbf {\bibinfo {volume} {16}},\ \bibinfo {pages} {7562} (\bibinfo {year} {2025})}\BibitemShut {NoStop}%
\bibitem [{\citenamefont {Ren}\ \emph {et~al.}(2025)\citenamefont {Ren}, \citenamefont {Zhang}, \citenamefont {Zeng}, \citenamefont {Zuo},\ and\ \citenamefont {Liu}}]{Ren2025JFMParticleJump}%
  \BibitemOpen
  \bibfield  {author} {\bibinfo {author} {\bibfnamefont {Z.}~\bibnamefont {Ren}}, \bibinfo {author} {\bibfnamefont {W.}~\bibnamefont {Zhang}}, \bibinfo {author} {\bibfnamefont {Q.}~\bibnamefont {Zeng}}, \bibinfo {author} {\bibfnamefont {Z.}~\bibnamefont {Zuo}},\ and\ \bibinfo {author} {\bibfnamefont {S.}~\bibnamefont {Liu}},\ }\href {https://doi.org/10.1017/jfm.2025.10998} {\bibfield  {journal} {\bibinfo  {journal} {Journal of Fluid Mechanics}\ }\textbf {\bibinfo {volume} {1025}},\ \bibinfo {pages} {A64} (\bibinfo {year} {2025})}\BibitemShut {NoStop}%
\bibitem [{\citenamefont {Ren}\ \emph {et~al.}(2022)\citenamefont {Ren}, \citenamefont {Zuo}, \citenamefont {Wu},\ and\ \citenamefont {Liu}}]{Ren2022}%
  \BibitemOpen
  \bibfield  {author} {\bibinfo {author} {\bibfnamefont {Z.}~\bibnamefont {Ren}}, \bibinfo {author} {\bibfnamefont {Z.}~\bibnamefont {Zuo}}, \bibinfo {author} {\bibfnamefont {S.}~\bibnamefont {Wu}},\ and\ \bibinfo {author} {\bibfnamefont {S.}~\bibnamefont {Liu}},\ }\href {https://doi.org/10.1103/PhysRevLett.128.044501} {\bibfield  {journal} {\bibinfo  {journal} {Physical Review Letters}\ }\textbf {\bibinfo {volume} {128}},\ \bibinfo {pages} {044501} (\bibinfo {year} {2022})}\BibitemShut {NoStop}%
\bibitem [{\citenamefont {Kling}\ and\ \citenamefont {Hammitt}(1972)}]{Kling1972}%
  \BibitemOpen
  \bibfield  {author} {\bibinfo {author} {\bibfnamefont {C.~L.}\ \bibnamefont {Kling}}\ and\ \bibinfo {author} {\bibfnamefont {F.~G.}\ \bibnamefont {Hammitt}},\ }\href {https://doi.org/10.1115/1.3425571} {\bibfield  {journal} {\bibinfo  {journal} {Journal of Basic Engineering}\ }\textbf {\bibinfo {volume} {94}},\ \bibinfo {pages} {825} (\bibinfo {year} {1972})}\BibitemShut {NoStop}%
\bibitem [{\citenamefont {Karri}\ \emph {et~al.}(2012)\citenamefont {Karri}, \citenamefont {Ohl}, \citenamefont {Klaseboer}, \citenamefont {Ohl},\ and\ \citenamefont {Khoo}}]{Karri2012}%
  \BibitemOpen
  \bibfield  {author} {\bibinfo {author} {\bibfnamefont {B.}~\bibnamefont {Karri}}, \bibinfo {author} {\bibfnamefont {S.-W.}\ \bibnamefont {Ohl}}, \bibinfo {author} {\bibfnamefont {E.}~\bibnamefont {Klaseboer}}, \bibinfo {author} {\bibfnamefont {C.-D.}\ \bibnamefont {Ohl}},\ and\ \bibinfo {author} {\bibfnamefont {B.~C.}\ \bibnamefont {Khoo}},\ }\href {https://doi.org/10.1103/PhysRevE.86.036309} {\bibfield  {journal} {\bibinfo  {journal} {Physical Review E}\ }\textbf {\bibinfo {volume} {86}},\ \bibinfo {pages} {036309} (\bibinfo {year} {2012})}\BibitemShut {NoStop}%
\bibitem [{\citenamefont {Teran}\ \emph {et~al.}(2018)\citenamefont {Teran}, \citenamefont {Rodriguez}, \citenamefont {La{\'i}n},\ and\ \citenamefont {Jung}}]{Teran2018}%
  \BibitemOpen
  \bibfield  {author} {\bibinfo {author} {\bibfnamefont {L.~A.}\ \bibnamefont {Teran}}, \bibinfo {author} {\bibfnamefont {S.~A.}\ \bibnamefont {Rodriguez}}, \bibinfo {author} {\bibfnamefont {S.}~\bibnamefont {La{\'i}n}},\ and\ \bibinfo {author} {\bibfnamefont {S.}~\bibnamefont {Jung}},\ }\href {https://doi.org/10.1063/1.5063472} {\bibfield  {journal} {\bibinfo  {journal} {Physics of Fluids}\ }\textbf {\bibinfo {volume} {30}},\ \bibinfo {pages} {123304} (\bibinfo {year} {2018})}\BibitemShut {NoStop}%
\bibitem [{\citenamefont {Huang}\ \emph {et~al.}(2024)\citenamefont {Huang}, \citenamefont {Ren}, \citenamefont {Liu},\ and\ \citenamefont {Hu}}]{Huang2024PoF}%
  \BibitemOpen
  \bibfield  {author} {\bibinfo {author} {\bibfnamefont {X.}~\bibnamefont {Huang}}, \bibinfo {author} {\bibfnamefont {C.-Z.}\ \bibnamefont {Ren}}, \bibinfo {author} {\bibfnamefont {P.-B.}\ \bibnamefont {Liu}},\ and\ \bibinfo {author} {\bibfnamefont {H.-B.}\ \bibnamefont {Hu}},\ }\href {https://doi.org/10.1063/5.0230263} {\bibfield  {journal} {\bibinfo  {journal} {Physics of Fluids}\ }\textbf {\bibinfo {volume} {36}},\ \bibinfo {pages} {103301} (\bibinfo {year} {2024})}\BibitemShut {NoStop}%
\bibitem [{\citenamefont {Lamb}(1932)}]{Lamb1932}%
  \BibitemOpen
  \bibfield  {author} {\bibinfo {author} {\bibfnamefont {H.}~\bibnamefont {Lamb}},\ }\href@noop {} {\emph {\bibinfo {title} {Hydrodynamics}}},\ \bibinfo {edition} {6th}\ ed.\ (\bibinfo  {publisher} {Cambridge University Press},\ \bibinfo {address} {Cambridge},\ \bibinfo {year} {1932})\BibitemShut {NoStop}%
\bibitem [{\citenamefont {Milne-Thomson}(1968)}]{MilneThomson1968}%
  \BibitemOpen
  \bibfield  {author} {\bibinfo {author} {\bibfnamefont {L.~M.}\ \bibnamefont {Milne-Thomson}},\ }\href@noop {} {\emph {\bibinfo {title} {Theoretical Hydrodynamics}}},\ \bibinfo {edition} {5th}\ ed.\ (\bibinfo  {publisher} {Macmillan},\ \bibinfo {address} {London},\ \bibinfo {year} {1968})\BibitemShut {NoStop}%
\bibitem [{\citenamefont {Brennen}(2014)}]{Brennen2014}%
  \BibitemOpen
  \bibfield  {author} {\bibinfo {author} {\bibfnamefont {C.~E.}\ \bibnamefont {Brennen}},\ }\href@noop {} {\emph {\bibinfo {title} {Cavitation and Bubble Dynamics}}}\ (\bibinfo  {publisher} {Cambridge University Press},\ \bibinfo {address} {Cambridge},\ \bibinfo {year} {2014})\BibitemShut {NoStop}%
\bibitem [{\citenamefont {Leighton}(1994)}]{Leighton1994}%
  \BibitemOpen
  \bibfield  {author} {\bibinfo {author} {\bibfnamefont {T.~G.}\ \bibnamefont {Leighton}},\ }\href@noop {} {\emph {\bibinfo {title} {The Acoustic Bubble}}}\ (\bibinfo  {publisher} {Academic Press},\ \bibinfo {address} {London},\ \bibinfo {year} {1994})\BibitemShut {NoStop}%
\bibitem [{\citenamefont {Zhang}\ \emph {et~al.}(2023)\citenamefont {Zhang}, \citenamefont {Li}, \citenamefont {Cui}, \citenamefont {Li},\ and\ \citenamefont {Liu}}]{Zhang2023Unified}%
  \BibitemOpen
  \bibfield  {author} {\bibinfo {author} {\bibfnamefont {A.-M.}\ \bibnamefont {Zhang}}, \bibinfo {author} {\bibfnamefont {S.-M.}\ \bibnamefont {Li}}, \bibinfo {author} {\bibfnamefont {P.}~\bibnamefont {Cui}}, \bibinfo {author} {\bibfnamefont {S.}~\bibnamefont {Li}},\ and\ \bibinfo {author} {\bibfnamefont {Y.-L.}\ \bibnamefont {Liu}},\ }\href {https://doi.org/10.1063/5.0145415} {\bibfield  {journal} {\bibinfo  {journal} {Physics of Fluids}\ }\textbf {\bibinfo {volume} {35}},\ \bibinfo {pages} {033323} (\bibinfo {year} {2023})}\BibitemShut {NoStop}%
\bibitem [{\citenamefont {Blake}\ \emph {et~al.}(2015)\citenamefont {Blake}, \citenamefont {Leppinen},\ and\ \citenamefont {Wang}}]{Blake2015}%
  \BibitemOpen
  \bibfield  {author} {\bibinfo {author} {\bibfnamefont {J.~R.}\ \bibnamefont {Blake}}, \bibinfo {author} {\bibfnamefont {D.~M.}\ \bibnamefont {Leppinen}},\ and\ \bibinfo {author} {\bibfnamefont {Q.}~\bibnamefont {Wang}},\ }\href {https://doi.org/10.1098/rsfs.2015.0017} {\bibfield  {journal} {\bibinfo  {journal} {Interface Focus}\ }\textbf {\bibinfo {volume} {5}},\ \bibinfo {pages} {20150017} (\bibinfo {year} {2015})}\BibitemShut {NoStop}%
\bibitem [{\citenamefont {Weiss}(1944)}]{Weiss1944}%
  \BibitemOpen
  \bibfield  {author} {\bibinfo {author} {\bibfnamefont {P.}~\bibnamefont {Weiss}},\ }\href {https://doi.org/10.1017/S0305004100018430} {\bibfield  {journal} {\bibinfo  {journal} {Proceedings of the Cambridge Philosophical Society}\ }\textbf {\bibinfo {volume} {40}},\ \bibinfo {pages} {259} (\bibinfo {year} {1944})}\BibitemShut {NoStop}%
\bibitem [{\citenamefont {Landweber}\ and\ \citenamefont {Miloh}(1980)}]{LandweberMiloh1980}%
  \BibitemOpen
  \bibfield  {author} {\bibinfo {author} {\bibfnamefont {L.}~\bibnamefont {Landweber}}\ and\ \bibinfo {author} {\bibfnamefont {T.}~\bibnamefont {Miloh}},\ }\href {https://doi.org/10.1017/S0022112080002005} {\bibfield  {journal} {\bibinfo  {journal} {Journal of Fluid Mechanics}\ }\textbf {\bibinfo {volume} {96}},\ \bibinfo {pages} {33} (\bibinfo {year} {1980})}\BibitemShut {NoStop}%
\bibitem [{\citenamefont {Wang}\ \emph {et~al.}(2022)\citenamefont {Wang}, \citenamefont {Wu}, \citenamefont {Zheng}, \citenamefont {Du}, \citenamefont {Zhang},\ and\ \citenamefont {Zhang}}]{Wang2022Weiss}%
  \BibitemOpen
  \bibfield  {author} {\bibinfo {author} {\bibfnamefont {X.}~\bibnamefont {Wang}}, \bibinfo {author} {\bibfnamefont {G.}~\bibnamefont {Wu}}, \bibinfo {author} {\bibfnamefont {X.}~\bibnamefont {Zheng}}, \bibinfo {author} {\bibfnamefont {X.}~\bibnamefont {Du}}, \bibinfo {author} {\bibfnamefont {Y.}~\bibnamefont {Zhang}},\ and\ \bibinfo {author} {\bibfnamefont {Y.}~\bibnamefont {Zhang}},\ }\href {https://doi.org/10.1016/j.ultsonch.2022.106130} {\bibfield  {journal} {\bibinfo  {journal} {Ultrasonics Sonochemistry}\ }\textbf {\bibinfo {volume} {89}},\ \bibinfo {pages} {106130} (\bibinfo {year} {2022})}\BibitemShut {NoStop}%
\bibitem [{\citenamefont {Zhang}\ \emph {et~al.}(1993)\citenamefont {Zhang}, \citenamefont {Duncan},\ and\ \citenamefont {Chahine}}]{Zhang1993}%
  \BibitemOpen
  \bibfield  {author} {\bibinfo {author} {\bibfnamefont {S.}~\bibnamefont {Zhang}}, \bibinfo {author} {\bibfnamefont {J.~H.}\ \bibnamefont {Duncan}},\ and\ \bibinfo {author} {\bibfnamefont {G.~L.}\ \bibnamefont {Chahine}},\ }\href {https://doi.org/10.1017/S0022112093003027} {\bibfield  {journal} {\bibinfo  {journal} {Journal of Fluid Mechanics}\ }\textbf {\bibinfo {volume} {257}},\ \bibinfo {pages} {147} (\bibinfo {year} {1993})}\BibitemShut {NoStop}%
\bibitem [{\citenamefont {Borkent}\ \emph {et~al.}(2008)\citenamefont {Borkent}, \citenamefont {Arora}, \citenamefont {Ohl}, \citenamefont {de~Jong}, \citenamefont {Versluis}, \citenamefont {Lohse}, \citenamefont {M{\o}rch}, \citenamefont {Klaseboer},\ and\ \citenamefont {Khoo}}]{Borkent2008}%
  \BibitemOpen
  \bibfield  {author} {\bibinfo {author} {\bibfnamefont {B.~M.}\ \bibnamefont {Borkent}}, \bibinfo {author} {\bibfnamefont {M.}~\bibnamefont {Arora}}, \bibinfo {author} {\bibfnamefont {C.-D.}\ \bibnamefont {Ohl}}, \bibinfo {author} {\bibfnamefont {N.}~\bibnamefont {de~Jong}}, \bibinfo {author} {\bibfnamefont {M.}~\bibnamefont {Versluis}}, \bibinfo {author} {\bibfnamefont {D.}~\bibnamefont {Lohse}}, \bibinfo {author} {\bibfnamefont {K.~A.}\ \bibnamefont {M{\o}rch}}, \bibinfo {author} {\bibfnamefont {E.}~\bibnamefont {Klaseboer}},\ and\ \bibinfo {author} {\bibfnamefont {B.~C.}\ \bibnamefont {Khoo}},\ }\href {https://doi.org/10.1017/S002211200800253X} {\bibfield  {journal} {\bibinfo  {journal} {Journal of Fluid Mechanics}\ }\textbf {\bibinfo {volume} {610}},\ \bibinfo {pages} {157} (\bibinfo {year} {2008})}\BibitemShut {NoStop}%
\bibitem [{\citenamefont {Li}\ \emph {et~al.}(2018)\citenamefont {Li}, \citenamefont {Zhang}, \citenamefont {Wang},\ and\ \citenamefont {Han}}]{Li2018PoF}%
  \BibitemOpen
  \bibfield  {author} {\bibinfo {author} {\bibfnamefont {S.}~\bibnamefont {Li}}, \bibinfo {author} {\bibfnamefont {A.-M.}\ \bibnamefont {Zhang}}, \bibinfo {author} {\bibfnamefont {S.-P.}\ \bibnamefont {Wang}},\ and\ \bibinfo {author} {\bibfnamefont {R.}~\bibnamefont {Han}},\ }\href {https://doi.org/10.1063/1.5044237} {\bibfield  {journal} {\bibinfo  {journal} {Physics of Fluids}\ }\textbf {\bibinfo {volume} {30}},\ \bibinfo {pages} {082111} (\bibinfo {year} {2018})}\BibitemShut {NoStop}%
\end{thebibliography}%

\end{document}


\title{Supplemental Material for: ``A Free Sphere Reverses the Rebound Direction of a Near-Wall Cavitation Bubble''}

\author{Chun-Zhu Ren}
\affiliation{School of Marine Science and Technology, Northwestern Polytechnical University, Xi'an, China}

\author{Jun Wen}
\affiliation{School of Marine Science and Technology, Northwestern Polytechnical University, Xi'an, China}

\author{Hai-Bao Hu}
\affiliation{School of Marine Science and Technology, Northwestern Polytechnical University, Xi'an, China}

\author{A-Man Zhang}
\affiliation{College of Shipbuilding Engineering, Harbin Engineering University, Harbin, China}

\author{Xiao Huang}
\email{huangxiao@nwpu.edu.cn}
\affiliation{School of Marine Science and Technology, Northwestern Polytechnical University, Xi'an, China}

\date{\today}
\maketitle

\section{Envelope-bubble radius history}

The potential-flow calculation requires a smooth envelope radius history, \(R_b(t)\). Therefore, the raw envelope radius extracted from the images is fitted using a Rayleigh--Plesset-type equation with a linear damping term:
\begin{equation}
    \dot{R}_b = U_b,
    \qquad
    \dot{U}_b = \frac{p_g(R_b) - p_\infty}{\rho R_b}
                - \frac{3}{2}\frac{U_b^2}{R_b} - \beta U_b.
    \label{eqS:rp_ode}
\end{equation}
Here, \(\rho\) is the liquid density, \(p_\infty\) is the far-field pressure, and \(\beta\) is a damping coefficient used solely to smooth the measured envelope history. The gas pressure used in the fit is given by
\begin{equation}
    p_g(R_b) = p_0\left(\frac{R_0}{R_b}\right)^{3\gamma} + P_0,
    \label{eqS:pg_law}
\end{equation}
where \(R_0\) and the starting time \(t_0\) are fixed, while \(p_0\), \(P_0\), \(\beta\), and \(\dot{R}_b(t_0)\) are optimized for each case. The fitted curve is constrained by the measured maximum radius, the time at which the maximum radius is reached, an intermediate measured point, and the first minimum collapse radius. We minimize the objective function
\begin{equation}
    \mathcal{E} =
    w_1\left[\frac{R_b(t_m) - R_{b,\max}}{R_{b,\max}}\right]^2
    + w_2\left[\frac{\dot{R}_b(t_m)}{U_{\mathrm{sc}}}\right]^2
    + w_3\left[\frac{R_b(t_s) - R_s}{R_s}\right]^2
    + w_4\left[\frac{R_b(t_e) - R_0}{R_0}\right]^2,
    \label{eqS:fit_objective}
\end{equation}
where \(t_m\) is the observed peak-radius time, \(t_s\) is an intermediate experimental time, \(t_e\) is the end of the fitted first cycle, and \(U_{\mathrm{sc}} = R_{b,\max}/[(t_e - t_0)/2]\) is used solely to make the velocity residual non-dimensional. Finally, the fitted radius provides the source strength used by the potential-flow model:
\begin{equation}
    Q(t) = 4\pi R_b^2 \dot{R}_b,
    \qquad
    \dot{Q}(t) = 4\pi\left(2R_b \dot{R}_b^2 + R_b^2 \ddot{R}_b\right).
    \label{eqS:Q}
\end{equation}

\section{Potential-flow calculation of the sphere motion}

\subsection{Geometry and velocity potential}

The bubble centre is fixed on the symmetry axis at \(z=z_b\).  The sphere centre
is \(z=z_p(t)\), and the rigid wall is represented by mirror singularities.  On
the sphere surface,
\begin{equation}
 x=R_p\sin\theta,
 \qquad
 z=z_p+R_p\cos\theta,
 \label{eqS:sphere_surface}
\end{equation}
where \(\theta=0\) is the upper pole of the sphere.  When the spherical envelope
of radius \(R_b\) intersects the sphere, the contact angle is
\begin{equation}
 \theta_c=
 \cos^{-1}\left(\frac{d^2+R_p^2-R_b^2}{2dR_p}\right),
 \qquad d=|z_b-z_p|,
 \label{eqS:theta_c}
\end{equation}
with \(\theta_c=0\) in non-contacting cases.  The liquid-wetted part of the
sphere is \(\theta\in[\theta_c,\pi]\); the contact cap is
\(\theta\in[0,\theta_c]\).

The liquid is taken as incompressible and irrotational outside the bubble and the
sphere.  The velocity potential is decomposed into four physical groups,

\begin{equation}
 \phi
 =
 \phi_{\rm bub}
 +
 \phi_{\rm ibub}
 +
 \phi_{\rm sph}
 +
 \phi_{\rm isph}.
 \label{eqS:phi_decomposition}
\end{equation}
The bubble source is
\begin{equation}
 \phi_{\rm bub}=\frac{Q}{4\pi r_b},
 \qquad
 r_b=|\bmx-z_b\bme_z|,
 \label{eqS:source_potential}
\end{equation}
and \(\phi_{\rm ibub}\) is the image source at \(-z_b\).  The sphere-induced
correction is represented by axial multipoles centred at the instantaneous sphere
centre,
\begin{align}
 \phi_{\rm sph}=&\frac{M_z}{4\pi}\frac{z-z_p}{r_p^3}
 +\frac{Q_z}{4\pi}\frac{3(z-z_p)^2-r_p^2}{2r_p^5}
 \nonumber\\
 &+\frac{O_z}{4\pi}
 \frac{(z-z_p)\left[5(z-z_p)^2-3r_p^2\right]}{2r_p^7},
 \qquad
 r_p=\left[x^2+(z-z_p)^2\right]^{1/2} .
 \label{eqS:multipole_potential}
\end{align}
The image-sphere potential \(\phi_{\rm isph}\) has the corresponding reflected
multipoles about \(-z_p\).  The coefficients \(M_z,Q_z,O_z\) are recalculated at
each time step from the instantaneous state \((Q,z_b,z_p,v_p)\) by imposing the
no-penetration condition on the sphere,

\begin{equation}
 \left.\frac{\partial \phi}{\partial n_p}\right|_{r_p=R_p}
 =
 -v_p\cos\theta .
 \label{eqS:nopenetration}
\end{equation}

\subsection{Non-contact pressure}

On the liquid-wetted part of the sphere, the pressure is evaluated from the
unsteady Bernoulli equation,
\begin{equation}
 p(\theta,t)
 =
 \rho\partial_t\phi
 -\frac{1}{2}\rho|\bm\nabla\phi|^2 .
 \label{eqS:bernoulli}
\end{equation}
The spatially uniform Bernoulli constant is omitted, since it cancels when the
pressure force is assembled over the closed sphere surface.

The non-contact vertical force is obtained by integrating the pressure over the
liquid-wetted part of the sphere,
\begin{equation}
 F_{\rm nc}
 =
 -2\pi R_p^2
 \int_{\theta_c}^{\pi}
 p(\theta,t)\sin\theta\cos\theta\,\diff\theta ,
 \label{eqS:F_nc}
\end{equation}
where \(\theta_c=0\) before physical contact and \(\theta_c>0\) after a contact
cap has formed.  The contact cap itself is excluded from this non-contact
integral and is treated by the contact-pressure closure described below.

For comparison with the contact contribution, the non-contact force is separated
into unsteady-pressure and dynamic-pressure parts,
\begin{equation}
 F_{\rm nc}=F_{\rm un}+F_{\rm dyn},
 \label{eqS:F_split}
\end{equation}
with
\begin{align}
 F_{\rm un}
 &=
 -2\pi\rho R_p^2
 \int_{\theta_c}^{\pi}
 \partial_t\phi\,
 \sin\theta\cos\theta\,\diff\theta ,
 \label{eqS:F_un}\\
 F_{\rm dyn}
 &=
 \pi\rho R_p^2
 \int_{\theta_c}^{\pi}
 |\bm{\nabla}\phi|^{2}
 \sin\theta\cos\theta\,\diff\theta .
 \label{eqS:F_dyn}
\end{align}

\subsection{Contact-pressure closure}

After the bubble comes into contact with the sphere, the velocity potential near
the contact region is distorted and is therefore not used to evaluate the local
dynamic pressure.  Instead, the contact pressure is completed by using an
effective bubble pressure defined from the Bernoulli pressure at the non-contact
free end of the bubble, together with a power-consistency condition.

The bubble volume used in the contact-power closure is not evaluated from the
spherical envelope volume alone.  The true gas volume is obtained by subtracting
the volume of the envelope sphere that overlaps with the solid sphere,
\begin{equation}
 V_{\rm true}(t)
 =
 \frac{4\pi}{3}R_b^3(t)
 -
 V_{\rm int}\!\left(R_b(t),R_p,d(t)\right),
 \qquad
 d(t)=|z_b-z_p(t)| ,
 \label{eqS:V_true}
\end{equation}
where \(V_{\rm int}\) is the geometric intersection volume between the spherical
bubble envelope and the solid sphere. Because the non-contact free end of the bubble envelope is prescribed by the
measured radius \(R_b(t)\), the effective bubble pressure is evaluated from the
Bernoulli pressure at that point.  With surface tension neglected,
\begin{equation}
 p_g(t)
 =
 \left[
 \rho\partial_t\phi
 -\frac{1}{2}\rho|\bm{\nabla}\phi|^2
 \right]_{\rm free\,end}.
 \label{eqS:pg_true}
\end{equation}
This definition keeps \(p_g\) and the free-surface pressure \(p_f\) below in the
same Bernoulli pressure gauge.

The mean pressure on the sphere contact cap, \(p_c(t)\), is determined by a
power-consistency condition between the gas work and the pressure work exerted
on the sphere surface.  The sphere surface is divided into the
liquid-wetted part \(S_p^{f}\) and the bubble-contact cap \(S_p^{c}\).  We write
\begin{equation}
 p_g(t)\frac{\diff V_{\rm true}}{\diff t}
 =
 \int_{S_p^{f}}p_f u_n\,\diff S
 +
 p_c(t)\int_{S_p^{c}}u_n\,\diff S .
 \label{eqS:power_consistency}
\end{equation}
Here \(p_f\) is the Bernoulli pressure acting on the liquid-wetted part of the
sphere surface, \(u_n\) is the normal velocity of the corresponding sphere
surface element measured along the gas-volume normal, and \(p_c\) is the mean
pressure over the sphere contact cap. Hence,
\begin{equation}
 p_c(t)
 =
 \frac{
 p_g(t)\,\diff V_{\rm true}/\diff t
 -
 \int_{S_p^f}p_f u_n\,\diff S
 }{
 \int_{S_p^{c}}u_n\,\diff S
 } .
 \label{eqS:pc_closure}
\end{equation}

The vertical contact force acting on the sphere is approximated by the mean
contact pressure over the projected contact area,
\begin{equation}
 F_c=-p_c\pi a_z^2 .
 \label{eqS:F_contact}
\end{equation}
Here \(a_z\) is the contact radius, and the sign is consistent with the vertical
force convention used in the sphere-motion equation.

\subsection{Motion update and neglected small forces}

The contact pressure is not treated as a dynamic-pressure contribution.  In the
sphere-motion update, the contact force is grouped with the unsteady-pressure
contribution,
\begin{equation}
 F_{\rm un}^{\rm tot}=F_{\rm un}+F_c ,
 \label{eqS:Fun_total}
\end{equation}
where \(F_{\rm un}\) is the non-contact unsteady-pressure force and \(F_c\) is
the mean contact-pressure force defined in Eq.~\eqref{eqS:F_contact}.  The
sphere motion is then advanced from
\begin{equation}
 m_p\frac{\diff v_p}{\diff t}
 =
 F_{\rm un}^{\rm tot}+F_{\rm dyn}+R_w,
 \qquad
 \frac{\diff z_p}{\diff t}=v_p .
 \label{eqS:motion_equation}
\end{equation}
Here \(R_w\) represents the unilateral constraint imposed by the rigid wall.
Before lift-off, the wall reaction balances the part of the hydrodynamic impulse
that would drive the sphere into the wall, so that the sphere remains at
\(z_p=R_p\) with zero downward velocity.  Once the accumulated hydrodynamic
impulse gives an upward velocity, the wall constraint is released and the sphere
is allowed to move upward.

Gravity, buoyancy, viscous drag, and direct surface-tension effects are not
included in Eq.~\eqref{eqS:motion_equation}.  Their relative importance was
estimated from the dimensionless groups listed in Table~\ref{tabS:dimensional}.
The large Reynolds and Weber numbers indicate that viscous and capillary effects
are small compared with the inertial pressure impulse associated with the bubble
motion, while the Froude number shows that gravity does not control the
first-cycle sphere acceleration.

\begin{table}[t]
\caption{Dimensionless groups used to assess the secondary forces.}
\label{tabS:dimensional}
\begin{ruledtabular}
\begin{tabular}{cccc}
Dimensionless group & Symbol & Definition & Value \\
\hline
Reynolds number &
\(Re\) &
\(Re=\rho U R_p/\mu\) &
\(\simeq 1.1\times10^{4}\) \\
Weber number &
\(We\) &
\(We=\rho U^{2}R_p/\sigma\) &
\(\simeq 2.8\times10^{2}\) \\
Froude number &
\(Fr\) &
\(Fr=U/\sqrt{gR_p}\) &
\(\simeq 9.0\) \\
Bond number &
\(Bo\) &
\(Bo=\rho gR_p^{2}/\sigma\) &
\(\simeq 3.4\)
\end{tabular}
\end{ruledtabular}
\end{table}

\section{Kelvin impulse on the closed bubble boundary}

For a closed bubble boundary \(S\), the Kelvin impulse is
\begin{equation}
I_K=-\rho\int_S\phi\,\bmn\,\diff S,
 \label{eqS:IK_def}
\end{equation}
where \(\bmn\) is the outward normal of the liquid domain.  Before contact,
\(S\) is the visible bubble surface.  After contact, the visible free surface is
open, and the bubble--sphere contact cap must be added to close the boundary,
\begin{equation}
 S=S_b^f\cup S_b^{c} .
 \label{eqS:Sclosed}
\end{equation}
The free part \(S_b^{f}\) has known potential and normal velocity from the analytical
potential-flow model.  The contact cap \(S_b^{c}\) has known normal velocity from the
rigid motion of the sphere, but its potential is unknown and is reconstructed by
a boundary-element method.

\subsection{Boundary-element equation and boundary conditions}

Let
\begin{equation}
 G(\bmx,\bmy)=\frac{1}{4\pi|\bmx-\bmy|}
 \label{eqS:green}
\end{equation}
be the free-space Green function.  The boundary integral equation for a point
\(\bmx_i\in S\) is written as
\begin{equation}
 c_i\phi(\bmx_i)+\int_S\phi(\bmy)
 \frac{\partial G(\bmx_i,\bmy)}{\partial n_y}\,\diff S_y
 =\int_S G(\bmx_i,\bmy)q(\bmy)\,\diff S_y,
 \label{eqS:BIE_continuous}
\end{equation}
where
\begin{equation}
 q=\frac{\partial\phi}{\partial n}
 \label{eqS:q_def}
\end{equation}
with the same normal convention as Eq.~\eqref{eqS:IK_def}.  After collocation on
triangular surface elements, Eq.~\eqref{eqS:BIE_continuous} becomes
\begin{equation}
 H\phi=Gq .
 \label{eqS:BIE_matrix}
\end{equation}
Here \(G_{ij}\) is the single-layer influence of element \(j\) on collocation
point \(i\), and \(H_{ij}\) is the corresponding double-layer influence including
the solid-angle and constant-potential consistency correction.  Splitting the
unknowns into free-surface and contact-surface parts gives
\begin{equation}
 H^f\phi_b^f+H^c\phi_b^c=G^fq^f+G^cq^c .
 \label{eqS:BIE_block}
\end{equation}
The imposed boundary data are
\begin{align}
 \phi_b^{f}&=\phi_{\rm bub}^{f}+\phi_{\rm ibub}^{f}
       +\phi_{\rm sph}^{f}+\phi_{\rm isph}^{f},
       && \bmx\in S_b^{f},
       \label{eqS:bc_free_phi}\\
 q^{f}&=-\dot R_b,
       && \bmx\in S_b^{f},
       \label{eqS:bc_free_q}\\
 q^{c}&=-\bm U_p\cdot\bmn_{c},
       \qquad \bm U_p=v_p\bme_z,
       && \bmx\in S_b^{c}.
       \label{eqS:bc_contact_q}
\end{align}

The minus signs in Eqs.~\eqref{eqS:bc_free_q} and \eqref{eqS:bc_contact_q}
follow from the velocity convention \(\bm u=-\bnabla\phi\) used in the numerical
implementation.  The contact potential \(\phi_b^c\) is the unknown.  Since both
\(\phi_b^f\) and \(q^f\) are retained on \(S_b^{f}\), Eq.~\eqref{eqS:BIE_block} is solved
as an overdetermined weighted least-squares problem,
\begin{equation}
 \phi_b^c=\arg\min_{\psi}
 \left\|W^{1/2}\left(H^c\psi-r\right)\right\|_2^2,
 \qquad
 r=G^fq^f+G^cq^c-H^f\phi_b^f,
 \label{eqS:BEM_LS}
\end{equation}
where \(W\) is a diagonal area-weighting matrix.  This reconstruction provides
\(\phi\) on the full closed boundary \(S_b^f\cup S_b^{c}\), so Eq.~\eqref{eqS:IK_def}
can be evaluated after contact.

\subsection{Physical component decomposition}

The same BEM matrices are reused to decompose the closed-boundary Kelvin impulse
into physical contributions.  On the free surface,
\begin{equation}
 \phi_b^{f}=
 \phi_{\rm bub}^{f}+\phi_{\rm ibub}^{f}
 +\phi_{\rm sph}^{f}+\phi_{\rm isph}^{f} .
 \label{eqS:free_decomp}
\end{equation}
For each component \(k\), the contact-cap potential \(\phi_k^{c}\) is obtained from
\begin{equation}
 H^{c}\phi_k^{c}\simeq G^{f}q_k^{f}-H^{f}\phi_k^{f},
 \qquad
 k\in\{\rm bub,ibub,sph,isph\} .
 \label{eqS:component_solve}
\end{equation}
The free-surface source flux is assigned to the bubble-source component, whereas
the wall-image and sphere-induced components carry no imposed free-surface flux
in the component solve.  The additional contribution from the contact-surface
velocity boundary condition is
\begin{equation}
 H^{c}\phi_{\rm vel}^{c}\simeq G^{c}q^{c} .
 \label{eqS:vel_component}
\end{equation}
The vertical contribution of each component is
\begin{equation}
 I_{k}= -\rho\left[
 \int_{S_b^{f}}\phi_k^{f} n_z\,\diff S+
 \int_{S_b^{c}}\phi_k^{c} n_z\,\diff S
 \right],
 \qquad
 k\in\{\rm bub,ibub,sph,isph,vel\},
 \label{eqS:component_impulse}
\end{equation}
and the contact-surface part alone is
\begin{equation}
 I_{k}^{c}= -\rho\int_{S_b^{c}}\phi_k^{c} n_z\,\diff S .
 \label{eqS:contact_impulse}
\end{equation}
These definitions are used for the decomposition shown in the main text.

\section{Scaling derivation of the three impulse contributions}

The main-text decomposition shows that the Kelvin-impulse bias is governed by
three leading contributions: the bubble-source contribution on the contact
closure surface, the wall-image-source contribution on the free interface, and
the quadrupolar part of the sphere-induced field.  The image-sphere and
velocity-boundary contributions remain secondary in the present parameter range
and are therefore not retained in the leading scaling balance.  In this section
we derive the three time-dependent scaling forms used in the main text, before
any late-cycle evaluation is made.

The bubble-source strength is defined from the envelope-radius history as
\begin{equation}
 Q_b(t)=4\pi R_b^2\dot R_b .
 \label{eqS:Qb_scaling_def}
\end{equation}
Thus \(Q_b>0\) during expansion and \(Q_b<0\) during collapse.  Upward, away
from the wall, is taken as the positive \(z\)-direction.

\subsection{Contact-source contribution}

The contact cap is not evaluated from the local dynamic pressure.  As described
above, its mean pressure is obtained from the effective bubble pressure through
the power-consistency condition.  For the present scaling estimate, we use the
inertial pressure scale implied by the inviscid Rayleigh--Plesset balance.  After
removing the gauge-dependent constant pressure and neglecting viscous and
capillary terms,
\begin{equation}
 \Delta p_c
 \sim
 \rho\left(R_b\ddot R_b+\frac{3}{2}\dot R_b^2\right).
 \label{eqS:pc_RP_scale}
\end{equation}
With
\begin{equation}
 U_b\sim |\dot R_b|\sim \frac{|Q_b|}{R_b^2},
 \label{eqS:Ub_scale}
\end{equation}
and the collapse-time estimate \(\ddot R_b\sim U_b^2/R_b\), Eq.~\eqref{eqS:pc_RP_scale}
gives
\begin{equation}
 \Delta p_c\sim \rho U_b^2 .
 \label{eqS:pc_scale}
\end{equation}
Over the bubble time scale \(\tau_b\sim R_b/U_b\), the corresponding contact
potential scale is
\begin{equation}
 \phi_{\rm bub}^c
 \sim
 \frac{\Delta p_c}{\rho}\tau_b
 \sim
 U_bR_b
 \sim
 \frac{|Q_b|}{R_b}.
 \label{eqS:phi_c_scale}
\end{equation}
Retaining the sign of the collapsing bubble source gives
\begin{equation}
 \phi_{\rm bub}^{c}\sim \frac{Q_b}{R_b}.
 \label{eqS:phi_bub_c_scale}
\end{equation}

The vertical projected area of the contact closure surface scales as \(a_z^2\).
Hence
\begin{equation}
 I_{\rm bub}^{c}
 =
 -\rho\int_{S_b^c}\phi_{\rm bub}^{c} n_z\,{\rm d}S
 \simeq
 -C_c\rho Q_b\frac{a_z^2}{R_b}.
 \label{eqS:I_bub_c_scaling}
\end{equation}
During collapse, \(Q_b<0\), so this term gives a positive, away-from-wall
contribution.  This is the contact-source scaling tested in the main text.

\subsection{Wall-image-source contribution}

The wall-image source is located below the rigid wall.  On the bubble free
interface, its potential can be expanded in powers of \(R_b/z_b\).  The
constant part of this expansion gives no contribution to the Kelvin-impulse
integral.  The first anisotropic part has the scale
\begin{equation}
 \phi_{\rm ibub}^{f}
 \sim
 Q_b\frac{R_b}{z_b^2}\cos\theta ,
 \label{eqS:phi_ibub_scale}
\end{equation}
where \(\theta\) is the polar angle on the bubble free interface.  Substitution
into the free-interface Kelvin-impulse integral gives
\begin{equation}
 I_{\rm ibub}^{f}
 =
 -\rho\int_{S_b^f}\phi^f_{\rm ibub} n_z\,{\rm d}S
 \simeq
 C_w\rho Q_b\frac{R_b^3}{z_b^2}.
 \label{eqS:I_ibub_f_scaling}
\end{equation}
This is the wall-image-source scaling used in the main text.  Since \(Q_b<0\)
during collapse, the fitted sign of \(C_w\) corresponds to a wallward
contribution.

\subsection{Quadrupolar sphere-induced contribution}

The multipole-resolved decomposition shows that the leading signed contribution
from the real sphere on the bubble free interface is quadrupolar.  We therefore
write the quadrupolar free-interface impulse as
\begin{equation}
 I_{\rm quad}^{f}
 =
 -\rho\int_{S_b^f}\phi^f_{\rm quad}n_z\,{\rm d}S .
 \label{eqS:Iquad_def}
\end{equation}

For a vertical quadrupole of strength \(\mathcal Q_z\), direct integration over
the bubble free interface gives
\begin{equation}
 I_{Q,z}^{f}
 =
 \frac{\rho\mathcal Q_z}{4}
 \frac{R_b^2}{d^3}
 \left[
 \frac{2R_b}{d}
 +
 \frac{P_Q(R_b,d,\alpha)}
 {d\left(d^2+R_b^2+2dR_b\cos\alpha\right)^{3/2}}
 \right],
 \label{eqS:I_quad_exact_general}
\end{equation}
where \(\alpha\) is the polar opening angle of the bubble free interface and
\begin{align}
P_Q(R_b,d,\alpha)
={}&
-2R_b^4
-6R_b^3d\cos\alpha
-3R_b^2d^2\cos^2\alpha
-3R_b^2d^2
\nonumber\\
&+
R_bd^3\cos^3\alpha
-3R_bd^3\cos\alpha
+d^4(\cos^2\alpha-1).
\label{eqS:PQ_def_general}
\end{align}
The geometric opening angle is taken as
\begin{equation}
\alpha =
\begin{cases}
\displaystyle
\cos^{-1}\!\left(
\frac{R_p^2-R_b^2-d^2}{2dR_b}
\right), & d<R_b+R_p,\\[2ex]
\pi, & d\ge R_b+R_p.
\end{cases}
\label{eqS:alpha_piecewise}
\end{equation}
The first line corresponds to a truncated free interface caused by
bubble--sphere contact, whereas the second line corresponds to a complete free
interface.

The no-penetration condition on the rigid sphere gives the quadrupolar response
scale
\begin{equation}
 \mathcal Q_z \sim Q_b\frac{R_p^5}{d^3},
 \label{eqS:Qz_scale}
\end{equation}
with numerical factors absorbed into the fitted coefficient.  Equations
\eqref{eqS:I_quad_exact_general}--\eqref{eqS:Qz_scale} can be written in the
dimensionless form
\begin{equation}
 I_{\rm quad}^{f}
 =
 C_{\rm quad}\rho Q_bR_p\,
 \mathcal T\left(\frac{R_b}{R_p},\frac{d}{R_p},\alpha\right),
 \label{eqS:Iquad_T_form}
\end{equation}
where \(\mathcal T\) is the dimensionless geometric kernel obtained from the
free-interface integral.

For contact or near-contact states, \(d<R_b+R_p\), the free interface is
truncated.  We introduce
\begin{equation}
 \chi=\frac{R_b}{R_p},
 \qquad
 \mu=\frac{d}{R_p}.
 \label{eqS:chi_mu_def}
\end{equation}
For the states used in the quadrupolar scaling test, the relevant ranges are
\[
0.495\lesssim \chi \lesssim 1.632,
\qquad
0.880\lesssim \mu \lesssim 3.464 .
\]
Substituting the contact-angle relation\(\cos\alpha=(R_p^2-R_b^2-d^2)/(2dR_b)\)
into the exact finite-opening kernel shows that the contact-state kernel can be
written, to leading order, as
\begin{equation}
\mathcal T(\chi,\mu,\alpha)
=
\frac{
\left[(1-\chi)^2-\mu^2\right]
\left[
2\chi\mu^2-\mu^4+O\!\left((1-\chi)^2\right)
\right]
}
{32\mu^7}.
\label{eqS:T_contact_expansion}
\end{equation}
In the present range, the finite-size correction \((1-\chi)^2\) is secondary
relative to the dominant separation term \(\mu^2\).  Neglecting this correction
gives
\begin{equation}
\mathcal T(\chi,\mu,\alpha)
\simeq
-\frac{1}{32}
\left(\frac{R_p}{d}\right)^3
\left[
\frac{2R_b}{R_p}
-
\left(\frac{d}{R_p}\right)^2
\right].
\label{eqS:T_contact_simplified}
\end{equation}
Therefore, for contact or near-contact states,

\begin{equation}
I_{\rm quad}^{f}
\simeq
-\frac{C_{\rm quad}}{32}\rho Q_b
\left(
\frac{2R_bR_p^3}{d^3}
-
\frac{R_p^2}{d}
\right),
\qquad d<R_b+R_p .
\label{eqS:I_quad_contact_expanded}
\end{equation}
This is the contact-state quadrupolar scaling used in the main-text impulse
balance.

For non-contact states, \(d\ge R_b+R_p\), the free interface is complete and
\(\alpha=\pi\).  In this limit the leading complete-surface quadrupolar scale is
\begin{equation}
 I_{\rm quad}^{f}
 \sim
 C_{\rm quad}\rho Q_b
 \frac{R_b^3R_p^5}{d^7},
 \qquad d\ge R_b+R_p .
 \label{eqS:I_quad_noncontact_scaling}
\end{equation}

\subsection{Time-resolved test of the quadrupolar scaling}

To test the quadrupolar scale before any late-cycle reduction, we define a
piecewise quadrupolar basis

\begin{equation}
I_{\rm quad}^{f}(t)
\simeq
\begin{cases}
\displaystyle
-\frac{C_{\rm quad}}{32}\rho Q_bR_p
\left(\frac{R_p}{d}\right)^3
\left[
\frac{2R_b}{R_p}
-\left(\frac{d}{R_p}\right)^2
\right],
& d<R_b+R_p,\\[3ex]
\displaystyle
C_{\rm quad}\rho Q_b
\frac{R_b^3R_p^5}{d^7},
& d\ge R_b+R_p .
\end{cases}
\label{eqS:I_quad_piecewise_scaling}
\end{equation}

\begin{figure*}[t]
\centering
\includegraphics[width=\textwidth]{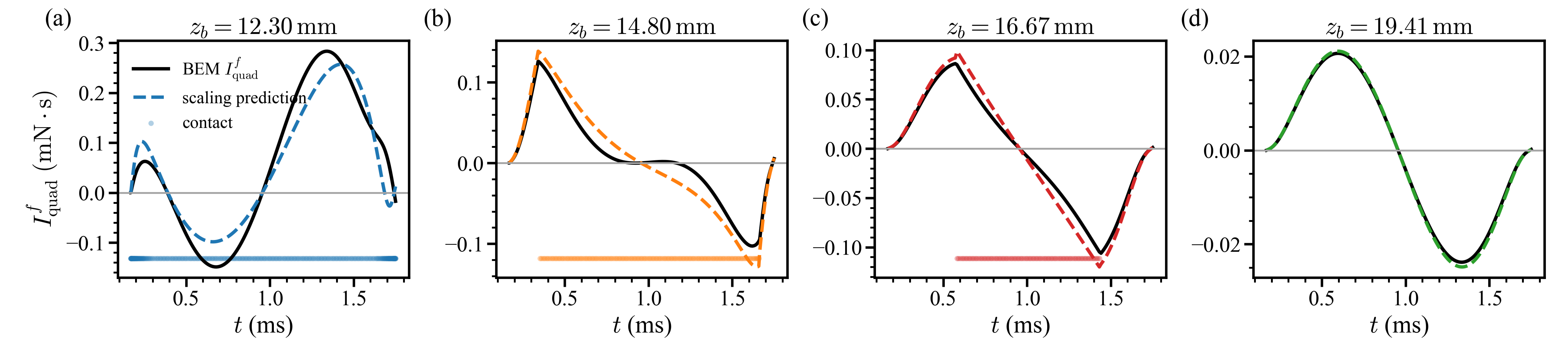}
\caption{
Time-resolved test of the quadrupolar scaling.
The BEM-resolved free-interface quadrupolar impulse \(I_{\rm quad}^{f}\) is
compared with the piecewise quadrupolar scaling prediction for
\(z_b=12.30\), \(14.80\), \(16.67\), and \(19.41\,\mathrm{mm}\).
The contact-state scaling is used when \(d<R_b+R_p\), whereas the
complete-free-surface scaling is used when \(d\ge R_b+R_p\).
}
\label{figS:quad_timeseries}

\end{figure*}

The BEM-resolved quadrupolar impulse is then compared with the piecewise
quadrupolar scaling prediction obtained from Eqs.~\eqref{eqS:T_contact_simplified}
and \eqref{eqS:I_quad_noncontact_scaling} for the four \(z_b\)-sweep cases shown
in Fig.~\ref{figS:quad_timeseries}.

